\documentclass[twocolumn]{aastex62}
\usepackage{amssymb}
\usepackage{array,multirow}
\usepackage{comment}
\usepackage{enumerate}
\usepackage{bm}

\newcommand{\msun}{{M_\odot}}

\newcommand{\gaia}{{\it Gaia}}

\newcommand{\Porb}{\ifmmode {P_{\rm orb}}\else${P_{\rm orb}}$\fi}

\newcommand{\Msun}{\ifmmode {{M_\odot}}\else{$M_\odot$}\fi}
\newcommand{\Mtot}{\ifmmode {{M_{\rm tot}}}\else{$M_{\rm tot}$}\fi}
\newcommand{\RV}{\ifmmode {{\rm RV}}\else RV \fi}
\newcommand{\bigG}{\ifmmode {\mathcal{G}}\else${\mathcal{G}}$\fi}

\newcommand{\emcee}{\texttt{emcee}}

\submitjournal{ApJ}

\shorttitle{Weighing the Darkness}
\shortauthors{Andrews, Breivik, and Chatterjee}

\begin{document}

\title{Weighing the Darkness: Astrometric Mass Measurement of Hidden Stellar Companions using \gaia}

\author[0000-0001-5261-3923]{Jeff J. Andrews}
\affiliation{Niels Bohr Institute, University of Copenhagen, Blegdamsvej 17, 2100 Copenhagen, Denmark}
\email{jeff.andrews@nbi.ku.dk}

\author[0000-0002-9660-9085]{Katelyn Breivik}
\affiliation{Canadian Institute for Theoretical Astrophysics, University
of Toronto, 60 St. George Street, Toronto, Ontario, M5S 1A7,
Canada}
\email{kbreivik@cita.utoronto.ca}

\author[0000-0002-3680-2684]{Sourav Chatterjee}
\affiliation{Tata Institute of Fundamental Research, Department of Astronomy and Astrophysics, Homi Bhabha Road, Navy Nagar, Colaba, Mumbai, 400005, India}
\email{souravchatterjee.tifr@gmail.com}

\begin{abstract}
In astrometric binaries, the presence of a dark, unseen star can be inferred from the gravitational pull it induces on its luminous binary companion. While the orbit of such binaries can be characterized with precise astrometric measurements, constraints made from astrometry alone are not enough to measure the component masses. In this work, we determine the precision with which \gaia\ can astrometrically measure the orbits and -- with additional observations -- the component masses, for luminous stars hosting hidden companions. Using realistic mock \gaia\ observations, we find that \gaia\ can precisely measure the orbits of binaries hosting hidden brown-dwarfs out to tens of pc and hidden white dwarf and neutron star companions at distances as far as several hundred pc. Heavier black hole companions may be measured out to 1 kpc or farther. We further determine how orbital period affects this precision, finding that \gaia\ can characterize orbits with periods as short as 10 days and as long as a few 10$^3$ days, with the best measured orbits having periods just short of \gaia's mission lifetime. Extending \gaia's nominal five-year mission lifetime by an additional five years not only allows for the measurement of longer period orbits, but those longer period binaries can be seen at even greater distances.
\end{abstract}

\keywords{black hole physics---methods: numerical---astrometry---binaries: general---stars: black holes}

\section{Introduction}
\label{S:intro}

The prospect of astrometrically detecting exoplanets through their gravitational pull on their host stars has long been lauded as one of \gaia's main contributions to exoplanet research \citep{Lattanzi2000, Sozzetti2005, Casertano2008, Perryman2014, Sozzetti2014, Ranalli2018}. Using \gaia's second data release \citep{Gaia_mission, Gaia_DR2_summary}, such predictions are beginning to be borne out with both the characterization of known exoplanets and the discovery of new exoplanet candidates \citep{Snellen2018, Livinston2018a, Livingston2018b, Bowler2018, Brandt2018, Brandt2018b, Dupuy2019}. 

However exoplanets are not the only objects to perturb the positions of stars; it has also been suggested that precise astrometric observations also have the potential to detect and characterize otherwise unseen black hole (BH) and neutron star (NS) companions to main sequence stars \citep{gould2002, Tomsick2010, Barstow2014}. Pioneering studies with ground-based interferometry have proven that precise tracking of the sky position of stars can afford some of the best measurements of binary star orbits, particularly when combined with spectroscopic radial velocities \citep{Hartkopf1996, Mason1999, Bowler2018, Gardner2018, Lucy2018}. Extensions to space-based observatories have so far been limited to a handful of binaries observed using the {\it Hubble Space Telescope}'s Fine Guidance Sensor \citep{Franz1998, Hershey1998, Benedict2000} and with {\it Hipparcos} \citep{Pourbaix2000, Halbwachs2000, Pourbaix2003, Goldin2007}.

Until recently, population studies of astrometric binary stars with \gaia\ focused on the millions of visual binaries comprised of pairs of main sequence stars \citep{Soderhjelm2004, Nurmi2005, Pourbaix2011}. However, two recent events -- occurring roughly simultaneously -- have led to a renewed focus on the astrometric characterization of compact object orbits. First, using binary population synthesis, several groups of researchers have shown that \gaia\ potentially has the ability to detect hundreds to thousands of detached compact object binaries \citep{Mashian2017, Breivik2017, Yamaguchi2018, Yalinewich2018}. Second, using the method first employed by \citet{Trimble1969}, multi-epoch spectroscopic campaigns have separately detected two binaries hosting detached compact object companions \citep{Giesers2018, Thompson2018}, thus confirming the existence of hidden companions that may show no observational signature apart from their gravitational effects on their luminous companions. In \citet{Breivik2018}, we show that improved mass constraints on one of these systems using \gaia\ astrometry will answer fundamental questions about BH formation.

\begin{figure}
    \begin{center}
    \includegraphics[width=1.0\columnwidth]{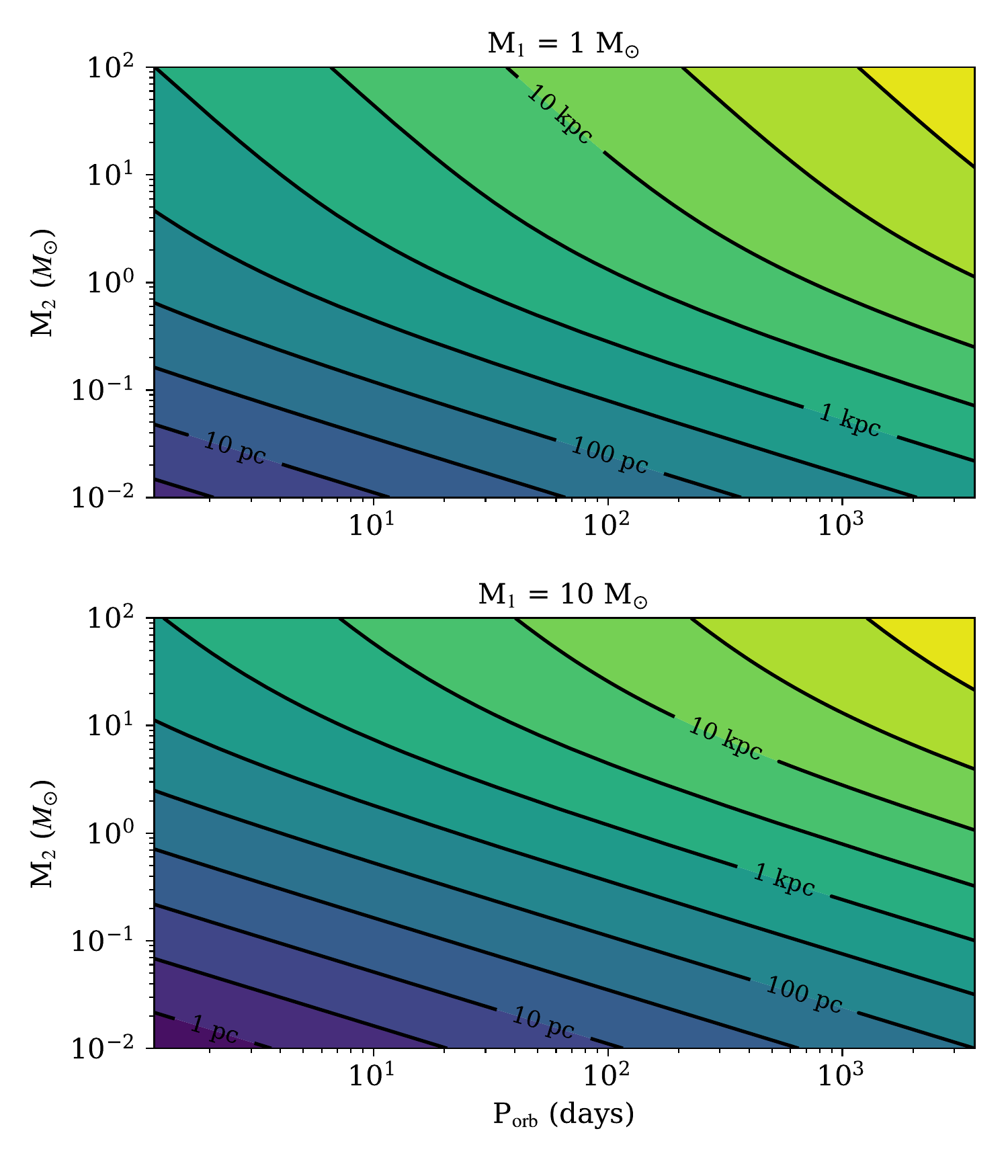}
    \caption{Using 0.1 mas as a fiducial estimate for the luminous star's orbital size that \gaia\ can accurately measure, we show the distances to binaries that can potentially be resolved. For \Porb $\sim$yrs, brown dwarfs are potentially detectable to dozens of pc, while massive compact object companions may be detectable at distances in excess of one kpc. }  
    \label{fig:horizon}
    
    \end{center}
\end{figure}

Anticipating the release of preliminary binary orbital determinations in the upcoming third \gaia\ data release, we wish to determine the types of orbits that \gaia\ will be able to measure. Adopting 0.1 mas as the fiducial size of the luminous star's orbit resolvable by \gaia, \autoref{fig:horizon} shows the ``horizon distance'' to binaries which can be astrometrically characterized for both 1 \Msun\ and 10 \Msun\ luminous star masses ($M_1$), as a function of their dark companion mass ($M_2$) and orbital period (\Porb). This figure indicates that \gaia\ may be able to characterize orbits for 10 \Msun\ BHs out to several kpc. 

Previous studies have demonstrated \gaia's ability to characterize planets \citep[e.g.,][]{Casertano2008}. With \autoref{fig:horizon} as motivation, we explore the precision with which \gaia\ can realistically characterize the masses of luminous stars and their hidden compact object and sub-stellar companions.  In \autoref{S:theory} we revisit relevant orbital dynamics calculations. Using modern statistical techniques and realistic mock \gaia\ observations, in \autoref{S:model} we describe our method for determining the precision with which \gaia\ will be able to measure binary star orbits. We provide the results of our models applied to luminous stars hosting hidden brown dwarf, white dwarf, NS, and BH companions in \autoref{S:results}. We discuss our results in \autoref{S:Discussion}, and finally we conclude in \autoref{S:Conclusion}.

\section{Mass Determination}
\label{S:theory}

The orbit of a star around a hidden companion is typically observed either spectroscopically (by measuring variations in a star's radial velocity) or astrometrically (by precisely mapping a star's changing position on the sky). Whereas radial velocity observations are limited by the unknown inclination angle (hence only a minimum mass of the unseen companion can be determined) astrometric observations provide a nearly complete orbital determination. However, even with exquisite data (using either or both of these methods), a degeneracy exists that does not allow the unique determination of the component masses of a binary. We review in turn these two orbital characterization methods and the resulting component mass degeneracy below. For a more complete derivation and exploration of binary orbits than what is provided below, we refer the reader to \citet{solar_system_dynamics}.

\subsection{Astrometry}

For binary star orbits, the orbital separation, $a$, the orbital period, \Porb, and the two stellar component masses, $M_1$ and $M_2$, are related by:
\begin{equation}
\left( \frac{2 \pi}{\Porb} \right)^2 = \frac{\bigG \Mtot }{a^3}, \label{eq:kepler_3}
\end{equation}
where $\Mtot = M_1+M_2$.

For a binary with an eccentricity $e$, the separation of the two components as a function of the true anomaly, $f$, is:
\begin{equation}
r = \frac{1-e^2}{1+e \cos f} a. \label{eq:orbital_sep}
\end{equation} 
If we want to split the orbital separation into two separations corresponding to the distance of each component to the binary's center of mass, we multiply the r.h.s.\ of \autoref{eq:orbital_sep} by a mass factor:
\begin{eqnarray}
r_1 &=& \frac{M_2}{\Mtot}\ \frac{1-e^2}{1+e \cos f}  a \nonumber \\
r_2 &=& \frac{M_1}{\Mtot}\ \frac{1-e^2}{1+e \cos f}  a. \label{eq:r_1}
\end{eqnarray}

All binaries suffer from perspective effects depending on the orientation of the binary relative to an observer. These are summarized by three angles: the inclination angle, $I$, the argument of periapse, $\omega$, and the longitude of the ascending node, $\Omega$. Dividing by the distance to the binary, $d$, provides an equation for perturbations to the right ascension, $\alpha$, and declination, $\delta$, of the stellar component of a binary relative to its center of mass:
\begin{eqnarray}
\Delta \alpha_1 &=& \frac{1}{\cos \delta}\frac{r_1}{d} \left[ \cos \Omega \cos (\omega+f) - \sin \Omega \sin(\omega+f) \cos I \right] \nonumber \\
\Delta \delta_1 &=& \frac{r_1}{d} \left[ \sin \Omega \cos (\omega+f) + \cos \Omega \sin(\omega+f) \cos I \right], \label{eq:astrometric_position}
\end{eqnarray}
where $r_1$ is derived from \autoref{eq:r_1}. An analogous equation (found by replacing $r_1$ with $-r_2$ in \autoref{eq:astrometric_position}) expresses the position of its dark companion. 

For a well-sampled orbit the angles, $f$, $\omega$, $\Omega$, and $I$, as well as the orbital parameters, $e$ and $P_{\rm orb}$, can %all (katie) 
be measured as precisely as the astrometric data allow. However, $a$ remains undetermined. Assuming the system's distance is known (for instance, using astrometric parallax), only $r_1$ can be measured observationally. Using \autoref{eq:kepler_3} and \autoref{eq:r_1} to substitute $a$ in favor of $r_1$, and using \autoref{eq:astrometric_position},
we find a relation between the unknown masses and the measured orbital parameters. For example, measured in terms of the orbital separation at apastron ($r_{\rm 1, ap}$), the masses are equal to:
\begin{equation}
\label{eq:astro_mf}
\frac{M_2}{\Mtot^{2/3}} = \frac{r_{\rm 1, ap}}{1 + e} \left( \frac{P_{\rm orb}}{2 \pi} \right)^{-2/3} \frac{1}{\mathcal{G}^{1/3}}.
\end{equation}
An analogous equation can be derived for the orbital separation at periastron. Regardless of the orbital phase, we can only obtain a measurement of the combined quantity $M_2/ \Mtot^{2/3}$. Additional information is required to break the degeneracy, which we discuss further in \autoref{S:mass_degeneracy}.

\subsection{Radial Velocities}

For completeness, we review the constraints on binary star masses from radial velocity measurements, demonstrating that these measurements also suffer from similar perspective effects as position. It can be shown that the radial velocity of a stellar component in a binary can be expressed as a function of $f$: 
\begin{eqnarray}
\Delta \RV_1 &=& \frac{M_2}{\Mtot} \frac{2\pi}{\Porb} \frac{a}{\sqrt{1-e^2}} \nonumber \\ 
 & & \quad \times\ \left[ \cos(\omega + f) \sin I + e \cos \omega \sin I \right]. \label{eq:rv_variation}
\end{eqnarray}
Subtracting the minimum radial velocity from the maximum radial velocity and dividing by two yields the radial velocity semi-amplitude, $K$:
\begin{equation}
K = \frac{M_2}{\Mtot^{2/3}} \left( \frac{\bigG 2 \pi}{\Porb}\right)^{1/3} \frac{1}{\sqrt{1-e^2}} \sin I. \label{eq:K}
\end{equation}

Assuming $K$, $e$, and $P_{\rm orb}$, are all well measured, \autoref{eq:K} produces a constraint on a term known as the mass function, $m_f$, which combines both component masses and $I$. This constraint parallels the constraint identified from astrometric observations:
\begin{eqnarray}
\label{eq:rv_mf}
m_f \equiv \frac{\left( M_2 \sin I \right)^3}{\Mtot^2} &=& \frac{K^3}{\bigG} \left( \frac{P_{\rm orb}}{2 \pi} \right) \left(1-e^2 \right)^{3/2}. \\
&=& \left(\frac{r_{\rm 1, ap}}{1 + e} \right)^3 \left( \frac{P_{\rm orb}}{2 \pi} \right)^{-2} \frac{\sin^3 I}{\mathcal{G}},
\end{eqnarray}
different only by the additional $\sin I$ degeneracy.

\subsection{Mass Degeneracy}
\label{S:mass_degeneracy}

With sufficiently precise data, an observer can determine the mass function either through radial velocity variations or astrometric observations (or both). Therefore, we can characterize the mass degeneracy in terms of $m_f$ and $I$.
Assuming astrometry provides measurements of both $m_f$ and $I$, the degeneracy between the component masses can be expressed as: 
\begin{equation}
    \frac{m_f}{\sin^3 I} = \frac{M_2^3}{\Mtot^2} = \frac{M_2^3}{\left(M_1 + M_2\right)^2} \label{eq:mass_degeneracy}
\end{equation}
Closed form solutions for the individual component masses exist. It is also worth noting that radial velocities add no new fundamental information to an astrometrically well-characterized binary star\footnote{Of course, RV measurements can immensely help characterize orbits with poor astrometric constraints.}. We return to this point in \autoref{S:Discussion}.

\begin{figure}
    \begin{center}
    \includegraphics[width=1.0\columnwidth]{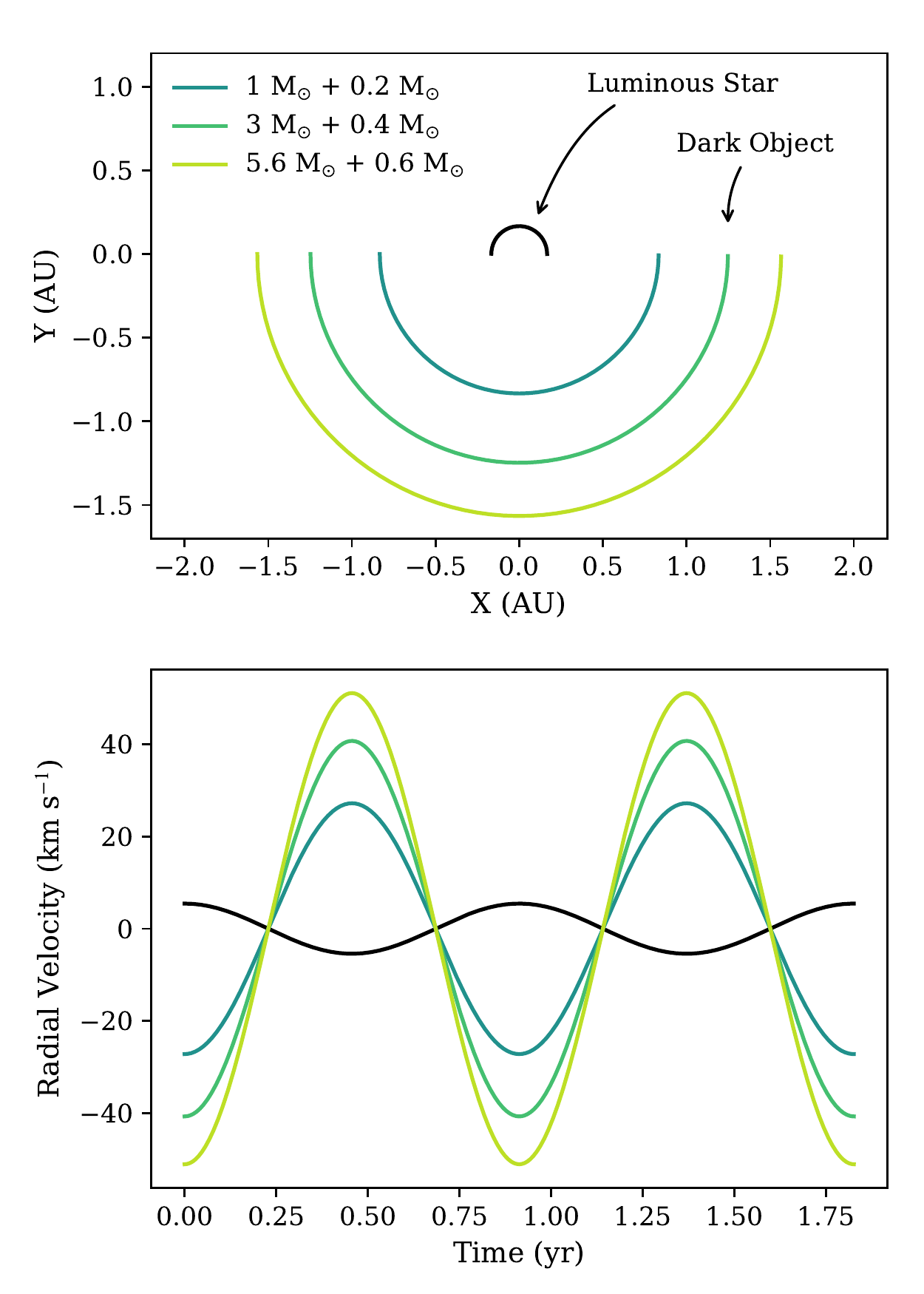}
    \caption{{\it Top:} Orbital integrations for half an orbit (hence the half-circles) show that identical astrometric data for the luminous star (black curve) can be formed by different possible combinations for the two stars' masses. Colored lines show the orbits of the unseen companions. {\it Bottom:} For the same combinations of masses with the same orbital periods (but over two complete orbits), the radial velocities of the luminous companion are also identical. Additional information, such as a mass estimate of the luminous star or radial velocity variations from its dark companion, are required to break the degeneracy.}  
    \label{fig:mass_degeneracy}
    
    \end{center}
\end{figure}

As demonstration, in \autoref{fig:mass_degeneracy} we numerically integrate the orbits of binary stars of different component masses. If we only see one star (black orbit), we cannot uniquely determine the mass combination for both stellar components; different combinations of the two component masses produce identical orbital positions (top panel) and radial velocities (bottom panel). The colored lines show the unseen orbits and radial velocities of the dark component.

\begin{figure}
    \begin{center}
    \includegraphics[width=1.0\columnwidth]{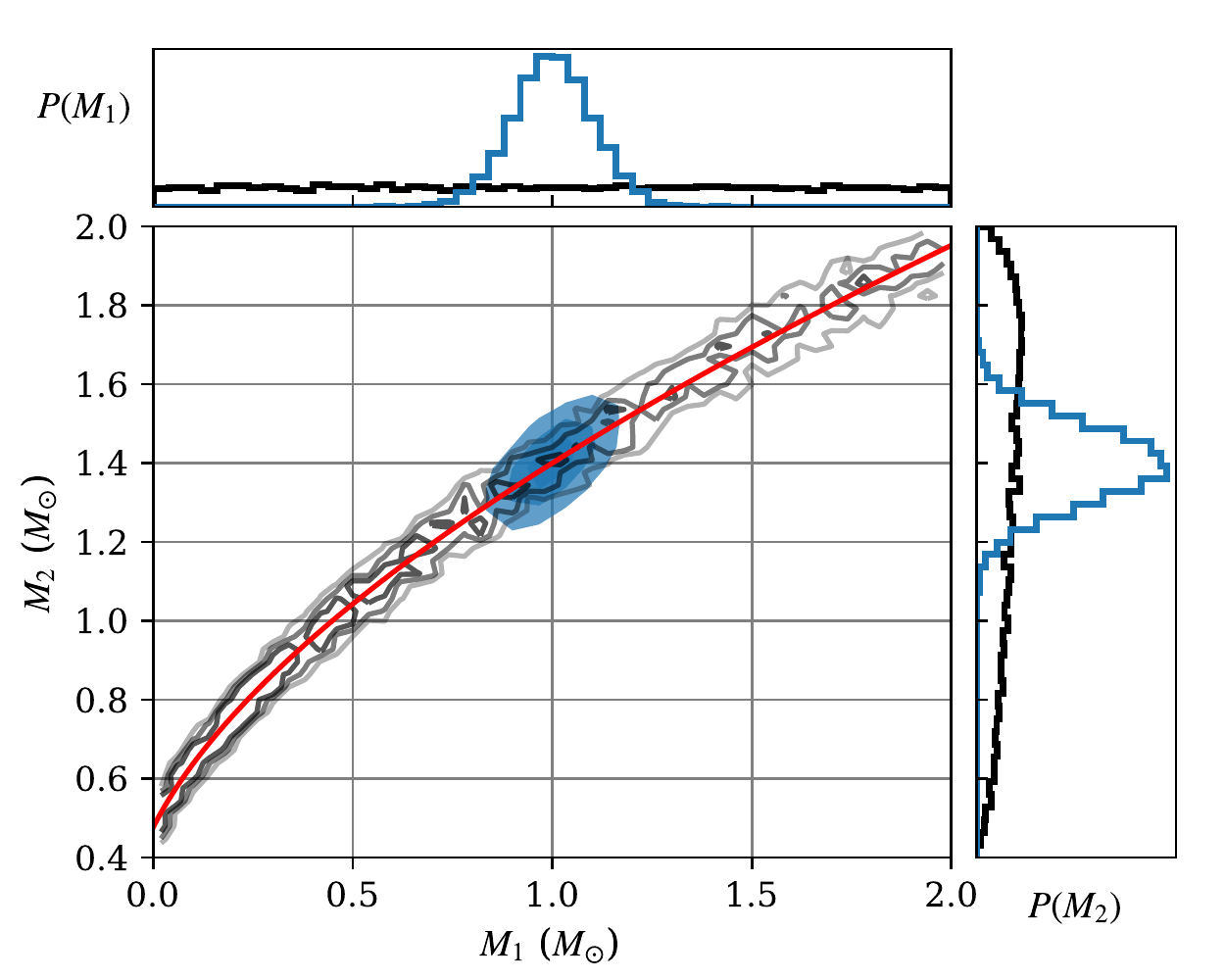}
    \caption{Using \autoref{eq:mass_degeneracy}, a 10\% measurement on the quantity $M_2^3/\Mtot^2$ for a 1 \Msun\ star with a 1.4 \Msun\ NS companion produces a joint constraint in $M_1$-$M_2$ space (black contours and lines). An additional 10\% measurement on the mass of the luminous star breaks the degeneracy, leading to a NS mass measurement within 0.1 \Msun\ (blue contours and lines).}
    \label{fig:mass_constraint}
    
    \end{center}
\end{figure}

Assuming $m_f$ and $I$ are well measured, an independent estimate of the luminous star's mass is sufficient to break this degeneracy. Typically, comparison to multi-band photometric predictions from stellar isochrones \citep[e.g., MIST;][]{Choi2016} can simultaneously provide estimates for the mass of and distance to the luminous component \citep{Kraus2007, Juric2008}. Alternatively, in close binaries the unseen companion may induce ellipsoidal variations as well as Doppler boosting, which can be observed and modeled with precise photometry \citep[e.g.,][]{Hermes2014}. Recently, \citet{Lin2018} and \citet{Mints2018} have shown how the inclusion of astrometric constraints from \gaia\ can provide improved photometric mass estimates. 

In \autoref{fig:mass_constraint}, we explicitly show the degeneracy between the two component masses described by \autoref{eq:mass_degeneracy} (black contours and lines) for a 10\% measurement on the quantity $M_2^3/\Mtot^2$ for a 1 \Msun\ star with a hidden 1.4 \Msun\ NS companion. Without extra information, a wide range of NS masses is possible\footnote{In \autoref{fig:mass_constraint} we randomly draw $M_1$ from a uniform distribution between 0 and 2 \Msun; values of $M_2$ are derived from these Monte Carlo draws of $M_1$, assuming a 10\% measurement precision on $M_2^3/\Mtot^2$.}. However, if a parallactic distance and an apparent magnitude can be compared with stellar models to obtain a 10\% estimate on $M_1$, the degeneracy is broken (blue contours and lines). For this combination of parameters, 10\% measurements on both $M_2^3/M_{\rm tot}^2$ and $M_1$ translate into a 0.1 \Msun\ precision (1-$\sigma$) mass measurement on the hidden NS.

\section{Statistical Model}
\label{S:model}

We wish to quantify \gaia's ability to characterize a binary's orbit for a broad range of component masses, distances, and orbital periods. \autoref{eq:astrometric_position} provides the aberration of a star's position due to a companion as a function of orbital parameters and $f$, which can be determined at a particular time by solving Kepler's equation. The apparent positions of stars are additionally perturbed by the effects of parallax and proper motion, each of which must be simultaneously modelled, lest incorrect orbital parameters are derived \citep[see discussion in][]{Torres2007}. We use the code {\tt SKYCALC} (J.\ Thorstensen, private communication) to calculate the motion of a star's position due to parallactic shifts ($\Delta \alpha_{\varpi}$, $\Delta \delta_{\varpi}$) as a function of the observation epoch and sky position. Finally, the proper motion ($\mu_{\alpha}$, $\mu_{\delta}$) is included as a term that increases linearly with time, so that for a given set of orbital parameters (including a reference epoch $\tau$), the position of the luminous star at a particular observation time ($t_{\rm obs}$) is deterministic:
\begin{eqnarray}
    \alpha_1 &=& \alpha + \Delta \alpha_1 + \Delta \alpha_{\varpi} + \mu_{\alpha}(t_{\rm obs} - \tau) \nonumber \\
    \delta_1 &=& \delta + \Delta \delta_1 + \Delta \delta_{\varpi} + \mu_{\delta}(t_{\rm obs} - \tau),
\end{eqnarray}
where the sky position ($\alpha$, $\delta$) is given for the center of mass of the system. Our generative model includes 13 parameters: $\alpha$, $\delta$, $M_1$, $M_2$, $\omega$, $\Omega$, $I$, $\tau$, \Porb, $e$, $\mu_{\alpha}$, $\mu_{\delta}$, and $d$, which we aggregate into a summarizing variable $\Theta$.

Using the \gaia\ observation forecast tool\footnote{https://gaia.esac.esa.int/gost/} we determine the epochs that \gaia\ will visit any particular field and the position angle with which \gaia's field of view transits. We randomly choose a field that has 75 separate observation times (roughly the sky average) over \gaia's nominal five-year mission lifetime. Using these observation times and values for the 13 parameters of any individual binary, we generate a time-series of synthetic sky positions for the luminous star. We explicitly vary $M_1$, $M_2$, \Porb, and $d$ for the synthetic binaries. We set $\tau$ to be the date of the first \gaia\ observation of these coordinates and keep the other parameters at fixed values: $\alpha=165.57283$, $\delta=41.22094$, $\Omega=\pi/3$, $\omega=\pi/3$, $I=\pi/3$, $e=0.01$, $\mu_{\alpha}=10$ mas yr$^{-1}$, and $\mu_{\delta}=10$ mas yr$^{-1}$. Although our results will vary somewhat depending on the exact choice of binary parameters, we opt to use the same parameter values throughout this study for consistent comparison between models.

Before describing in detail our procedure for producing synthetic \gaia\ observations, we review some of the peculiarities of \gaia\ data. Since it is a scanning mission, the position of every star is measured separately along two axes, one along the direction of a star's motion across \gaia's field of view (``along scan"; AL) and the other perpendicular to its motion (``across scan"; AC). In the AL direction \gaia\ can measure the position of a star to $\sim$1 mas, with the exact precision depending on the star's magnitude. For stars with $G<13$, this precision can improve to $<$0.1 mas, while for fainter stars with $G=21$, this precision reduces to $\approx$10 mas \citep{Gaia_astrometry}. In the AC direction, \gaia's astrometric measurement precision is currently 612 mas; the error ellipses describing the sky position of a star observed at a single epoch are extremely narrow and long, with axis ratios of $\sim$100-to-1 or greater\footnote{Note that we take these numbers by analyzing the position error ellipses from the released \gaia\ epoch astrometry of Solar System objects \citet{Gaia_solar_system}. Future calibrations ought to reduce astrometric uncertainties in the AC direction, so the error ellipses will ultimately have axis ratios of $\sim$10 \citep{lindegren12}. However, even with these improvements, measurements in the AC direction minimially aid in the overall astrometric constraints.}. Since the spacecraft precesses as it orbits the Earth and the Earth orbits the Sun, every time \gaia\ observes a field, the rotation angle varies between the AL-AC and $\alpha$-$\delta$ coordinate frames. Fits to the error ellipses defining a set of observations of a star, each with its own position angle, produce the parallax and proper motion parameter estimates within the \gaia\ catalogs.

To simulate realistic \gaia\ observational errors, we separately apply random perturbations in the observed position for the AL and AC directions. In the AC direction we apply offsets randomly drawn from a Gaussian with a standard deviation of 612 mas, while in the AL direction, we use a standard deviation corresponding to the single observation pointing precision ($\sigma_{\xi}$; note that this parameter is different from a star's parallax precision). For a star of a given magnitude, we determine $\sigma_{\xi}$ from fitting formula estimating \gaia's astrometric measurement precision \citep[formula provided by B.\ Holl, private communication; these are consistent with the red curve in Figure 9 from][]{Gaia_astrometry}\footnote{We use the calculated apparent magnitude of the luminous star placed at the distance to the nominal binary. We opt to ignore extinction for generality. However, note that extinction becomes relevant for stars at larger distances as well as low Galactic latitudes.}.

After separately applying the AL and AC measurement errors, we obtain a set of observed positions ($\alpha_{\rm obs}$, $\delta_{\rm obs}$) for each observation time, which we represent as a set of position vectors, $\bm{\mu_{\rm obs}}$.

We rotate a diagonal matrix containing the AL and AC measurement variances by the position angle, provided by the \gaia\ observation forecast tool, for each observation to find a set of covariance matrices ($\bm{\Sigma_{\rm obs}}$), one for each mock \gaia\ observation, describing the two-dimensional positional error region in the sky position. The astrometric precision in the AL direction is two orders of magnitude smaller than in the AC direction. Since the position angle that \gaia\ observes for each star rotates over time, the orbit is resolved even with a poor AC measurement precision.

Given a set of synthetic observations ($t_{\rm obs}$, $\bm{\mu_{\rm obs}}$,  $\bm{\Sigma_{\rm obs}}$), Bayes' theorem provides the posterior probability of a particular set of model parameters:
\begin{equation}
    P(\Theta | t_{\rm obs}, \bm{\mu_{\rm obs}}, \bm{\Sigma_{\rm obs}}) \sim P(\Theta) P(\bm{\mu_{\rm obs}}, \bm{\Sigma_{\rm obs}} | \Theta, t_{\rm obs}).
\end{equation}
We opt for flat priors on all model parameters except $I$, for which $P(I)=0.5 \sin I$ to account for an isotropic distribution of viewing angles, and $d$ for which we apply the exponentially decreasing space density prior from \citet{Astraatmadja2016}: $P(d)=(d^2/2 L^3) \exp(-d/L)$, where we set $L$ to 20 kpc (rather than their recommended value of 1 kpc) so as not to strongly bias our results for test binaries at distances similar to $L$. We can then calculate the log of the likelihood for a particular set of model parameters, $\Theta$:
\begin{eqnarray}
    &\ln& P(\bm{\mu_{\rm obs}}, \bm{\Sigma_{\rm obs}} | \Theta, t_{\rm obs}) = \nonumber \\
    && \qquad \sum_i^{75} - \frac{1}{2} \left[\bm{\mu_{\rm obs,i}}-\bm{\mu_{\Theta}}(t_i)\right]^{\rm T} \times \nonumber\\
    && \qquad \quad \quad \bm{\Sigma_{\rm obs}}^{-1} \left[\bm{\mu_{\rm obs,i}}-\bm{\mu_{\Theta}}(t_i)\right] \label{eq:likelihood}
\end{eqnarray}
where $\bm{\mu_{\Theta}}(t)$ is the position vector predicted by a model with parameters $\Theta$ at a time $t$.  

Using the priors and likelihood equation above, we can now determine how well a set of synthetic observations constrain the 13 model parameters characterizing the binary. While analysis using numerically calculated Fisher matrices can provide a quick, first-order solution, the non-linear nature of the posterior space, as well as the degeneracy between $M_1$ and $M_2$ leads us to use the Markov-Chain Monte Carlo sampler \emcee\ \citep{Foreman-Mackey2013} to explore the parameter space. 

Since we are interested in determining \gaia's ability to measure orbits, we avoid the issue of parameter initialization. For astrometric orbits, this can be a particularly thorny issue which has been addressed elsewhere in the literature \citep{Casertano2008, Schulze-Hartung2012, Lucy2014}. We return to the issue of parameter initialization in \autoref{S:Discussion}. As it is an ensemble sampler, \emcee\ simultaneously runs multiple walkers. We opt for 128 walkers, which we randomly initialize in a 13-dimensional ``ball'' around the true values of the parameters used to generate our synthetic data. The samplers are run for 40,000 steps, the first 10,000 of which are thrown away as the burn-in, and our results are produced from the remaining 30,000 samples. In what follows we will derive our results using specific example choices of binary parameters, however, this general scheme is applicable for any astrometric binary.

\section{Results}
\label{S:results}

\begin{figure*}
    \begin{center}
    \includegraphics[width=1.0\textwidth]{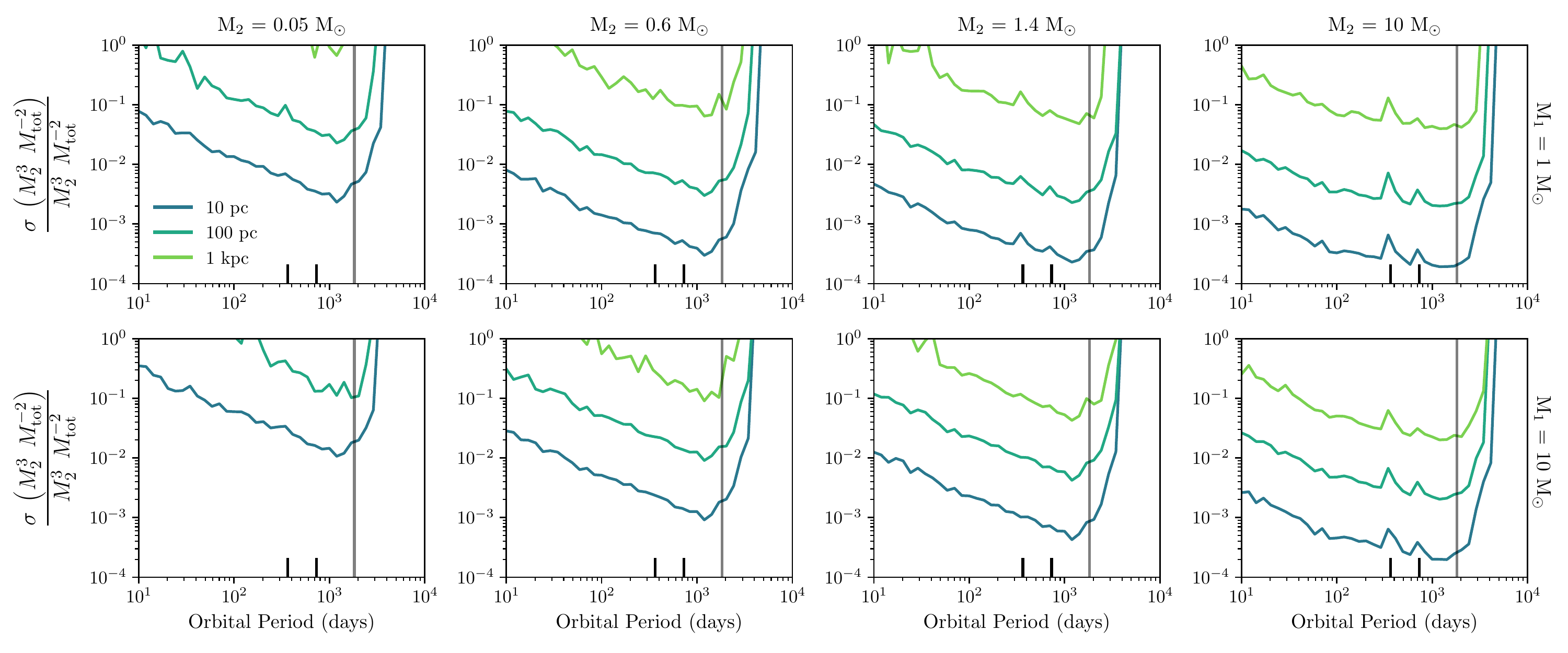}
    \caption{Each panel shows the relative precision with which \gaia\ can measure the combined quantity $M_2^3/M_{\rm tot}^2$ for different distances as a function of \Porb. The top and bottom rows of panels show constraints for luminous stars of 1 \Msun\ and 10 \Msun, respectively. Different columns explore the constraints for different mass hidden companions, corresponding to brown dwarfs (left), white dwarfs (left-center), NSs (right-center), and BHs (right). \gaia's ability to characterize orbits improves with increasing \Porb, up to the five-year mission lifetime (indicated by the vertical, black line). Degraded performance for orbital periods around one and two years (small ticks on the $x$-axis) is seen for hidden BHs and, to a lesser extent, for hidden NSs. }  
    \label{fig:model_constraints}
    
    \end{center}
\end{figure*}

To test \gaia's ability to measure orbits for a combination of binary types, we create a grid of models, varying the two component masses, the distance to the binary, and the binary's orbital period. We choose luminous stars with masses of either 1 or 10 \Msun, each with dark companions with masses corresponding to a brown dwarf ($M_2=0.05$ \Msun), white dwarf ($M_2=0.6$ \Msun), neutron star ($M_2=1.4$ \Msun), or black hole ($M_2=10$ \Msun). For each pair of possible component masses, we place the binary at three different distances: 10 pc, 100 pc, and 1 kpc. Finally, for each combination of component masses and distance, we explore orbital periods ranging from 10 to 10$^4$ days.

Due to the mass degeneracy described in \autoref{S:theory}, rather than attempting to constrain either component's mass directly, we will determine \gaia's astrometric ability to measure the combined quantity $m_f / \sin^3I$, which translates to $M_2^3/M_{\rm tot}^2$ using \autoref{eq:mass_degeneracy}. This measurement precision can be combined with estimates on the mass of the luminous star by some other means, to determine the measurement precision of the dark object.

In \autoref{fig:model_constraints}, we provide the model results on the relative error on the quantity $M_2^3/M_{\rm tot}^2$. These are obtained by calculating the quantity $M_2^3/M_{\rm tot}^2$ for the posterior samples for each model and taking half of the 1$\sigma$ contour intervals on this quantity around the median. 

Each separate panel shows results for different combinations of the component masses as a function of binary period; more massive dark objects cause larger effects on their luminous companions and therefore tend to be better measured. The different lines in each panel show the results for the same binary placed at different distances; since the projected orbital size scales inversely with distance, \gaia's ability to measure orbits degrades linearly with distance. For 1 \Msun\ luminous stars, the degradation is somewhat worse than linear at distances beyond 100 pc since \gaia's astrometric precision diminishes for more distant, and therefore fainter, stars. As expected, all of our models show that longer period (and therefore physically larger) orbits are better characterized than short orbital period ones. The vertical black lines in each panel show the nominal five-year mission lifetime of \gaia. Orbits with periods slightly less than this lifetime are best characterized, while \gaia's ability to characterize orbits quickly degrades for binaries with periods longer than \gaia's nominal observation duration (see also, Holl et al.\ 2019, in preparation). This suggests that extending \gaia's lifetime to ten years would have a profound effect on its ability to astrometrically characterize stellar binaries. We discuss this further in \autoref{S:rv_10yr}.

The right two columns show that BH-mass companions, and to a lesser extent NS-mass companions, exhibit a decreased sensitivity around orbital periods of a year and two years (marked with black ticks at the bottom of each panel). This is due to a partial degeneracy between parallax and orbital motion close to these periods \citep[see also][]{Butkevich2018}. 

We now separately describe our model results for each type of hidden companion.

\subsection{Brown Dwarfs}

The left column shows that substellar objects with masses $=0.05$ \Msun\ in binary systems and with 10$^2 < \Porb/{\rm days} < 10^3$ are detectable by \gaia\ within the nearest 10 pc. For brown dwarfs in $\sim$10$^3$ day orbits around solar-type stars, this may extend to nearly 100 pc. Previous studies have shown the benefit that astrometry can bring to characterizing brown dwarf binary systems \citep[][Holl et al.\ 2019, in preparation]{Halbwachs2000, Sahlmann2013, Bowler2018, Dieterich2018}. 

Estimates suggest that the Milky Way contains as many as 10$^{11}$ brown dwarfs \citep{Muzik2017}, with $\sim$10$^3$ brown dwarfs in the nearest 25 pc \citep{bardalez2019}. Since as many as 5\% of these brown dwarfs may be found in binaries ranging from 1-1000 au \citep{Fontanive2018}, we can predict that \gaia\ will characterize tens of such brown dwarf binaries. Many of these would normally go undetected by magnitude limited surveys such as \gaia; as demonstrated by \citet{Faherty2018}, standard methods of detecting brown dwarfs in \gaia\ data suffer in crowded, severely reddened fields. Since many of these brown dwarfs host significantly brighter companions, otherwise undetectable brown dwarfs may be identified at larger distances using the astrometric wobble they impose. Such searches are subject to fundamentally different selection effects and offer new opportunities to dynamically measure the masses of otherwise difficult-to-detect brown dwarfs.

\subsection{White Dwarfs}

Although their temperatures can be in excess of 10$^4$K, white dwarfs have radii a factor of 10$^3$ smaller than main sequence stars of similar mass. Some white dwarfs with binary companions, particularly those that are old with lower $T_{\rm eff}$, are effectively dark to most surveys, since they contribute only a small fraction of the overall luminosity of the system. The recent detection of a hidden white dwarf companion to a $\approx$1 \Msun\ star in a $\sim$1 year orbit confirms the existence of such systems \citep{Masuda2019}. 

\autoref{fig:model_constraints} shows that these systems can be astrometrically identified at distances greater than 100 pc. For orbital periods of several years, this distance may extend to nearly a kpc. This potentially provides accurate orbital parameters for symbiotic-type binaries. However, note that special care may need to be taken, since the white dwarfs' flux contributions in symbiotic binaries may be non-negligible, particularly when the system is not in quiescence. In certain cases, neither star may dominate the system's luminosity, and the motion of the photocenter of the system must be modeled, substantially complicating astrometric orbital derivations \citep{Coughlin2010, shahaf2019}.

\gaia's completeness for directly detecting WDs decreases at distances beyond 20 pc \citep{Carrasco2014}. Since \autoref{fig:model_constraints} shows that astrometric binaries may be detectable at somewhat larger distances, white dwarfs may be indirectly detected when found in astrometric binaries at distances outside of \gaia's nominal completeness volume. In addition, since \gaia\ detections are limited only by the orbit size and the magnitude of the bright companion, these detections will potentially be complimentary to what \gaia\  can otherwise detect. Such observations will provide critical opportunities to study the formation and evolution of WD binaries and their progenitors \citep{Toonen2017}.

\begin{figure*}
    \begin{center}
    \includegraphics[width=1.0\textwidth]{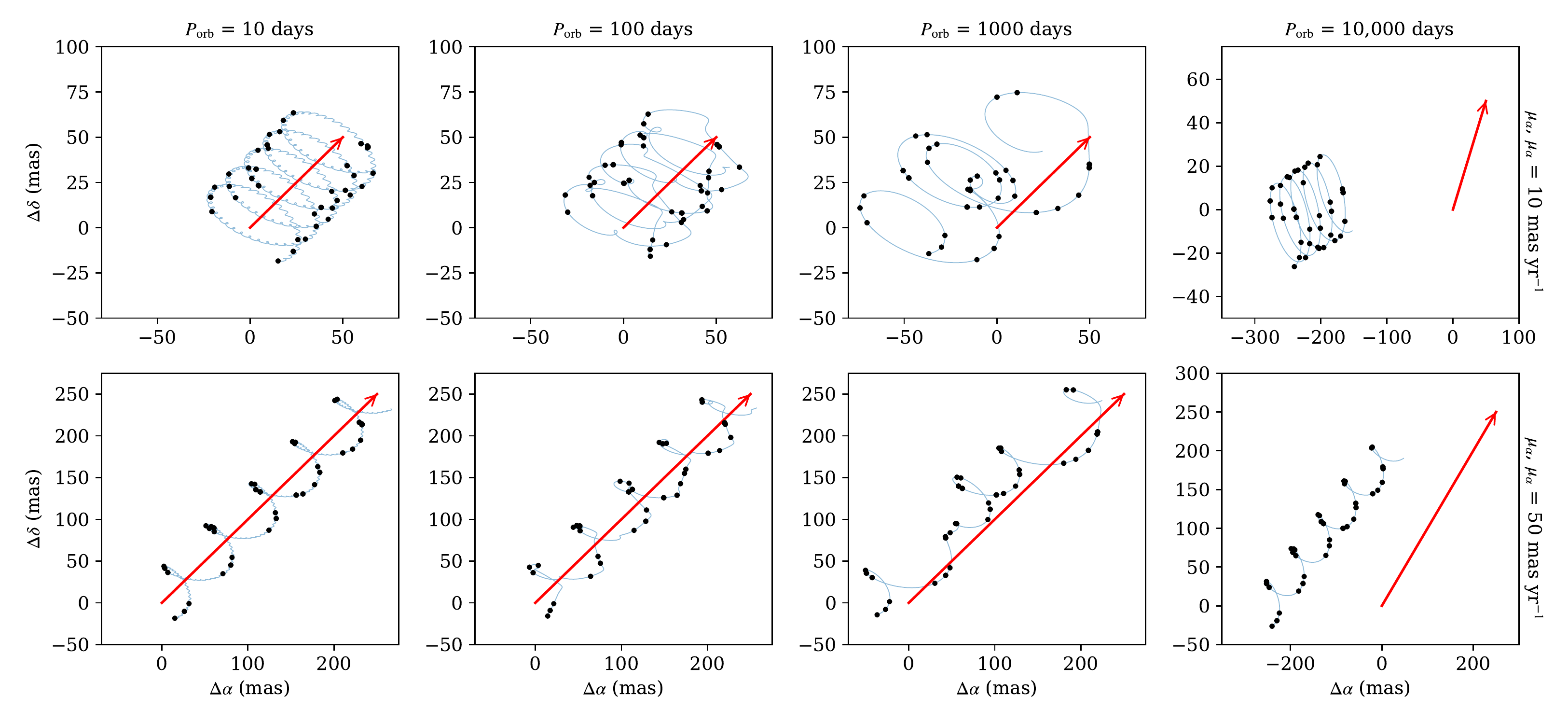}
    \caption{Each panel shows the astrometric motion of a $1\,\msun$ luminous star orbiting a $1.4\, \msun$ hidden companion with varying orbital periods and proper motions. For short period binaries, the motion of a star on the sky is replaced by a small helix. For binaries with \Porb$\approx$1 year, that motion becomes much more complex. At still longer \Porb, the parallax and orbital motions are no longer confused, but such orbital motion may be difficult to detect as any orbital motion may appear to be part of a star's proper motion vector (red arrow). }  
    \label{fig:parallax_orbit}
    
    \end{center}
\end{figure*}

\subsection{Neutron Stars}

With masses typically $\approx$1.4\Msun, NSs can cause a profound effect on the positions of stellar companions. The third column in \autoref{fig:model_constraints} shows that such systems may be measurable out to $\sim$1 kpc, for optimal \Porb. This precision greatly improves at close distances, such that wide NS binaries, if they exist within the nearest 100 pc, may allow $M_2^3/M_{\rm tot}^2$ to be measured to 1 per cent. Even at 1 kpc, $M_2^3/M_{\rm tot}^2$ may be measured to 10 per cent. \autoref{fig:mass_degeneracy} shows that, when combined with a 10 per cent measurement of the luminous companion, the masses of such NSs may be measured with a precision of 0.1 \Msun.

\subsection{Black Holes}

Currently the best way to identify (X-ray- and radio-faint) black hole companions to luminous stars is through radial velocity monitoring of a large number of stars \citep{Trimble1969}. Although studies employing this method have recently yielded new detections \citep{Giesers2018, Thompson2018}, the fourth column in \autoref{fig:model_constraints} suggests that \gaia\ will be able to characterize every black hole with a luminous companion over a wide range in \Porb\ within nearly one kpc with relative errors of 10 per cent or better. Recent population synthesis studies have shown such systems are ubiquitous throughout the Milky Way \citep[e.g.,][]{Breivik2017}. 
Furthermore, the astrometric detection and characterization of black hole binaries will be a great benefit to studies that depend on understanding black hole formation properties (e.g., mass function at birth), to make theoretical predictions for binary black hole merger rates and the properties expected to be observed by gravitational-wave or X-ray observatories \citep[e.g.,][]{Chatterjee2017,Breivik2018}.

\section{Discussion}
\label{S:Discussion}

\subsection{Analytic Estimates}

Examination of \autoref{fig:model_constraints} shows a self-similarity among the various curves. Based on our results across all our simulations, we find an analytic estimate for the relative measurement precision on $M_2^3/M_{\rm tot}^2$:
\begin{equation}
    \frac{\sigma \left(M_2^3\ M_{\rm tot}^{-2} \right)}{M_2^3\ M_{\rm tot}^{-2}}
    \approx 0.9 \left( \frac{\sigma_{\xi}}{\theta} \right)  \left( \frac{N}{75} \right)^{-1/2}, \enskip P_{\rm orb} < 5\ {\rm years} , \label{eq:self_similar}
\end{equation}
where $N$ is the number of \gaia\ observations of an orbit and $\theta \sim (a/d)\times M_2/M_{\rm{tot}}$ is the angular size of the orbit made by the luminous star on the sky plane. \autoref{eq:self_similar} is applicable for orbits with periods less than \gaia's mission lifetime and may be somewhat underestimated for orbits with periods close to one or two years. 

We apply this estimate to the simulated population of BHs and NSs orbiting giant stars (GSs) that are observable by \gaia\ in \citet{Breivik2018}. We find that, for an average of $75$
observations, the total number of systems that have mass measurements of $\sigma \left(M_2^3\ M_{\rm tot}^{-2} \right)/\left(M_2^3\ M_{\rm tot}^{-2}\right) < 0.1$ is 45$\pm$7 BH-GS binaries and 90$\pm$9 NS-GS binaries (see \autoref{tab:my_label}). We note that this is fewer by a factor of $2$ compared to the numbers deemed observable by \gaia\ in \citet{Breivik2018}. This is because in that earlier work we used an older, more optimistic estimate of the astrometric precision of \gaia. Nevertheless, even with the more rigorous analysis of \gaia's sensitivity, the expected yield comfortably surpasses the total number of compact objects with giant star companions found to date using methods relying on X-ray, radio, and radial velocity observations.

\begin{table}[]
    \centering
    \begin{tabular}{c|c|c|c}
        Source & \citep{Breivik2018} & $10\%$ error & $5\%$ error \\
        \tableline
        BH--GS & $74\pm9$ & $45\pm7$ & $20\pm4$\\
        NS--GS & $190\pm14$ & $90\pm9$ & $27\pm4$\\
    \end{tabular}
    \caption{Comparison of the number of BHs and NSs orbiting GSs from the Delayed model of \cite{Breivik2018} with $\theta > 3\sigma_{\xi}$ and the number of sources with relative mass measurement errors less than $10\%$ and $5\%$.}
    \label{tab:my_label}
\end{table}

\begin{figure*}
    \begin{center}
    \includegraphics[width=1.0\textwidth]{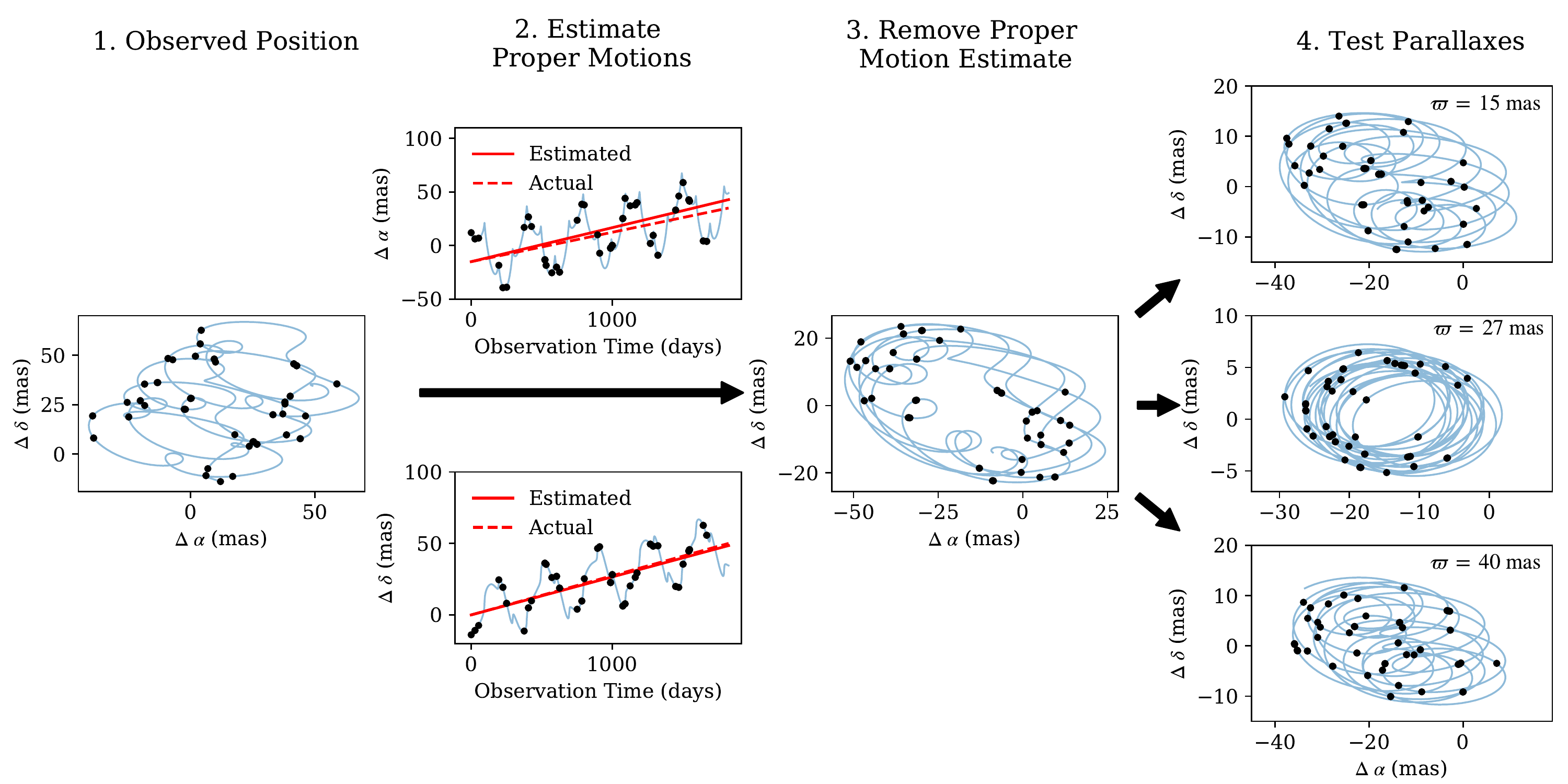}
    \caption{A proposed procedure to separate orbital motion from astrometric parallax and proper motion for a 1 \Msun\ star with a hidden 1.4 \Msun\ NS companion in a 100-day orbit at a distance of 35 pc ($\varpi=28.8$ mas). Motion of the luminous star is shown in blue, while a set of 75 \gaia\ observations over its five-year lifetime are shown as black points. The left panel shows the position on the sky. Separate linear fits to the right ascension and declination over time (second column of panels) provide estimates on the system's proper motion. Subtracting these estimates from the sky position produces a more ordered orbit (center-right panel). Different trial values of the parallax can be tested (right column of panels) until one is found that produces a near-elliptical orbit. The adopted proper motion and parallax values are sufficiently accurate to initialize the MCMC walkers. }  
    \label{fig:initialization}
    
    \end{center}
\end{figure*}

\subsection{Detection and Initialization}

Our results suggest \gaia\ ought to be able to characterize a wide variety of binary orbits containing dark objects. This exercise depends on the previous identification through some other analysis \citep[e.g., using \gaia's excess astrometric noise parameter;][]{evans2018} that a particular star actually contains a hidden companion. In this work, we ignore the issue of binary detection altogether, which is complicated by both the size of the \gaia\ dataset as well as the possibility of false detection. Although \gaia's detection efficiency has been previously discussed in the context of exoplanets \citep{Casertano2008, Ranalli2018} and brown dwarfs (Holl et al.\ 2019, in preparation), future work ought to extend these injection-detection studies to characterization of higher-mass hidden objects. 

Furthermore, in this work we consider only the precision with which \gaia\ can measure astrometric orbits. We initialize the orbits near the maximum in posterior space, ignoring the complexities of parameter initialization. Depending on the orbital parameters, this may be a particularly challenging task for future work, since parallax and proper motion must simultaneously be accounted for. \autoref{fig:parallax_orbit} shows the position of a 1 \Msun\ star orbiting a 1.4 \Msun\ dark companion over the course of five years; red arrows show movement of the binary's center of mass due to the system's proper motion. The binary is placed at a relatively close distance of 35 pc to magnify any astrometric effects. Black dots show the positions of the star over 75 epochs characterizing a typical non-uniform \gaia\ observation cadence. For orbital periods much less than a year, the orbital motion and parallax can be disentangled. However, for $\Porb =$ 10$^2$--10$^3$ days, the positions become extremely complex. 

The right two columns of \autoref{fig:parallax_orbit} show that at orbital periods longer than the mission lifetime, the binary orbital motion may be confused with proper motion rather than the perturbing effects from parallax \citep{Wielen1997}. Clearly orbits with periods somewhat longer than \gaia's lifetime will be challenging to detect and characterize, although this is actively being studied \citep{Lucy2014, Bowler2018, Docobo2018, Kervella2019a, Kervella2019b}.

How is one to initialize the walkers in an MCMC-like analysis that allows for the global maximum in posterior space to be identified in our 13-dimensional parameter space? Comparison between the top and bottom panels of \autoref{fig:parallax_orbit} shows that it may be easier to separate the effects of parallax and orbital motion in binaries with relatively larger proper motions. 

In \autoref{fig:initialization} we demonstrate an initialization procedure for the orbit of a 1 \Msun\ star with a hidden 1.4 \Msun\ NS companion in a 100-day orbit at a distance of 35 pc (the same orbit as shown in the top, middle-left panel of \autoref{fig:parallax_orbit}). Blue lines show the position of the luminous star, while the black points indicate the positions at 75 separate epochs at a typical irregular \gaia\ cadence. The second column of panels in \autoref{fig:initialization} separately shows the evolution of the star's position in $\alpha$ and $\delta$ as a function of time. Linear fits to these positions (red, solid lines) provide reasonably accurate estimates of the star's actual proper motion (red, dashed lines). Subtracting the effect of proper motion from the sky position of the star produces the distribution of position seen in the third column in \autoref{fig:initialization}. Trial parallaxes can then be subtracted to find the best possible closed orbits; the center right panel (with $\varpi=27$ mas) shows clear orbital motion, whereas both smaller and larger trial parallaxes produce unordered motion on the sky. Once a reasonably close trial parallax has been found, initialization of the remaining orbital parameters can proceed using methods described elsewhere \citep{Lucy2014}. Note that this procedure is merely meant to demonstrate that parallactic and orbital motions can be disentangled even for complex sky positions. In practice, more robust methods are used by the \gaia\ team (B.\ Holl, private communication). 

For neutron stars and black holes, astrometry -- if precise enough -- provides perhaps the most straightforward observations for deriving constraints on masses; such mass constraints may be the only way to determine the nature of the dark companion to a radial velocity variable, particularly if the dark companions are radio and X-ray dark. Furthermore, population synthesis studies \citep[e.g., ][]{Breivik2017} indicate the Galaxy ought to be flush with such long-period binaries. Radial velocity follow-up to identify these is both observationally expensive and technically demanding. For example, the on-going campaign to spectroscopically characterized long-period binary subdwarf B stars \citep{Vos2012, Vos2017, Vos2019} exhibits the challenges faced in long-term radial velocity campaigns.

\subsection{Radial Velocities and an Extended \gaia\ Mission}
\label{S:rv_10yr}

For distant or short-period binaries, the angular size of a binary's orbit becomes small compared with \gaia's angular resolution. In such cases, which are at the limit of \gaia's capabilities, astrometry provides only weak constraints on the orbital parameters (including the inclination angle). Follow-up observations using high-resolution spectroscopy may yield the precise radial velocities needed to aid astrometric measurements. However, for many binaries spectroscopic follow-up may not be required. Over its five-year nominal lifetime, \gaia\ takes an average of 40 separate radial velocity measurements of every star with $G\lesssim$16 using its radial velocity spectrometer \citep[although note that only the integrated spectra will be useful for stars this faint; epoch radial velocities with reasonable signal-to-noise require stars somewhat brighter;][]{Gaia_radial_velocities}. The precision of these radial velocity observations depend on both a given star's magnitude and its stellar type (late-type stars, which have more spectral lines, tend to be better measured), but may be $<$1 km s$^{-1}$.

To test the improvement in orbital determination provided by the inclusion of \gaia\ radial velocities, we extend our model described in \autoref{S:model} to include radial velocity variations as calculated using \autoref{eq:rv_variation}. Mock observation times are determined from the \gaia\ observation forecast tool, and we select only those times in which the star crosses \gaia's focal plane on CCDs 4-7, corresponding to the coverage by the radial velocity spectrometer. We simulate errors in the radial velocity measurements by adding Gaussian noise with a standard deviation ($\sigma_{\rm RV}$) of 1 km s$^{-1}$ to every mock observation. We then add the radial velocity term to the log-likelihood function in \autoref{eq:likelihood}: 
\begin{equation}
\ln P(\bm{{\rm RV}_{\rm obs}} | \Theta, t_{\rm obs}) = \sum_i^{40} - \frac{\left[{\rm RV}_{{\rm obs},i} - {\rm RV}_{\Theta}(t_i)\right]^2}{2\sigma_{\rm RV}^2},
\end{equation}
where ${\rm RV}_{{\rm obs},i}$ is the $i$th observed radial velocity and ${\rm RV}_{\Theta}(t_i)$ is the calculated radial velocity at time $t_i$ for the set of model parameters $\Theta$.

\begin{figure}
    \begin{center}
    \includegraphics[width=1.0\columnwidth]{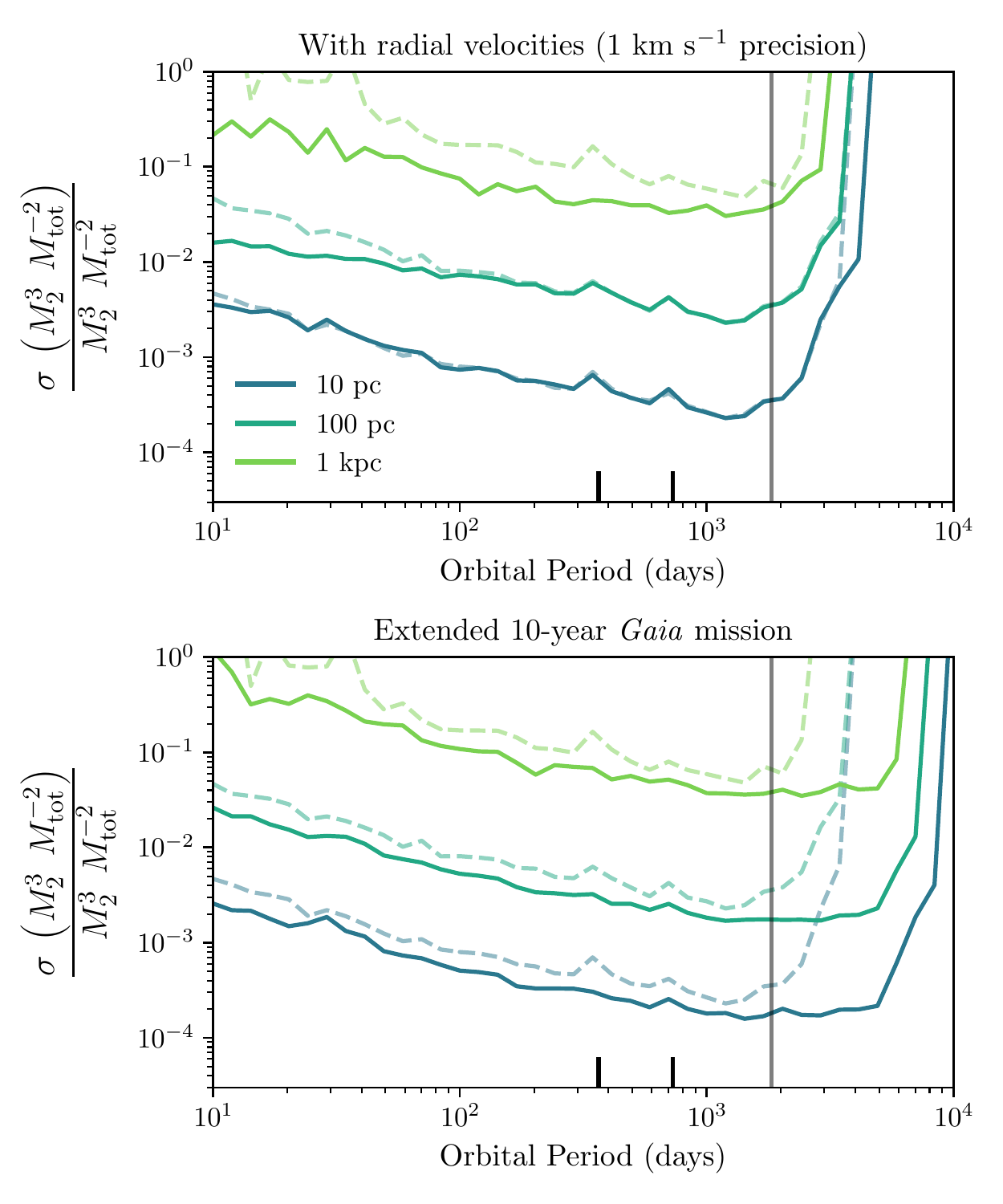}
    \caption{ Constraints the orbit of a binary comprised of a dark 1.4 \Msun\ NS companion to a 1 \Msun\ luminous star when \gaia\ radial velocities are included (top panel) and for astrometry only, but with an extended 10-year \gaia\ mission (bottom panel). Dashed lines show the constraints using astrometry only for a nominal, 5-year \gaia\ mission. See \autoref{S:rv_10yr} for details. The inclusion of radial velocities improves \gaia's sensitivity for binaries with short periods and large distances, whereas an extended, 10-year \gaia\ mission allows for binaries to be detected at much longer orbital periods. }
    \label{fig:RV_10yr}
    
    \end{center}
\end{figure}

The top panel of \autoref{fig:RV_10yr} compares the constraints on a binary's orbit using astrometry only (dashed lines) with the improved constraints that include \gaia\ radial velocities (solid lines). These constraints were produced for a dark 1.4 \Msun\ neutron star companion to a 1 \Msun\ luminous companion, with orbital parameters and simulation settings otherwise identical to those used to produce the constraints displayed in \autoref{fig:model_constraints}. For binaries at a distance of 10 pc, radial velocities - at least with 1 km s$^{-1}$ precision - negligibly aid in measuring the orbit. At larger distances and shorter orbital periods, radial velocities begin to substantially improve orbital determinations. However, note that \gaia's radial velocity precision is a strong function of magnitude. Ignoring extinction, a 1 \Msun\ star at a distance of 1 kpc has a \gaia\ $G\approx14.7$, far too faint to obtain epoch radial velocity measurements with \gaia's radial velocity spectrometer \citep{Gaia_radial_velocities}. Even at a distance of 100 pc, a 1 \Msun\ star is likely to have epoch radial velocity measurement precisions a factor of a few larger than the 1 km s$^{-1}$ precision used to produce the constraints in \autoref{fig:RV_10yr}. Therefore the potential gain from the inclusion of radial velocity measurements will go largely unrealized without ground-based radial velocity follow-up observations. 

In the bottom panel of \autoref{fig:RV_10yr}, we compare the astrometry-only constraints on the same binary (1.4 \Msun\ NS with a 1 \Msun\ luminous companion) for the nominal 5-year \gaia\ mission (dashed lines) with analogous constraints for an extended 10-year \gaia\ mission (solid lines). Doubling the mission lifetime leads to two separate effects on the orbital constraints. First, the measurement precision improves by a factor $\approx \sqrt{2}$ since the number of individual observations doubles. Second, \gaia\ becomes sensitive to binaries with orbital periods between five and ten years. Given the expansion to the range of orbital periods to which \gaia\ is sensitive, an extension to \gaia's mission lifetime would be a great benefit to the study of stellar binaries and exoplanets.

\section{Conclusion}
\label{S:Conclusion}

In this work we focus on the ability for \gaia\ to astrometrically measure the orbits of luminous stars with dark companions. We show that from astrometry and radial velocities alone, a degeneracy exists between the two stars' masses; this degeneracy can be broken if extra information, such as the mass of the luminous star, is provided.

Furthermore, we find that when parallax and proper motions are included, the motion of a star with a hidden companion on the sky becomes complex, particularly for orbits with periods similar to a year. Depending on the system parameters, this may be very challenging for \gaia\ to interpret.

Ignoring the issues of detection and initialization, we develop a statistical method, using Markov-Chain Monte Carlo, to determine how precise an orbit can be astrometrically measured. Applying this method to realistic mock \gaia\ observations that account separately for both the along-scan and across-scan position uncertainties, we quantify the distance to which \gaia\ can astrometrically characterize orbits for a 1 \Msun\ luminous star: tens of pc for brown dwarfs and hundreds of pc for hidden white dwarfs, neutron stars or black holes. As they are somewhat brighter, and hence have improved angular precision, hidden companions around more massive, 10 \Msun\ luminous stars may be better characterized, allowing NS and BH companions to be measured out to 1 kpc or more. It is therefore likely that \gaia\ will not only detect many new astrometric binaries with dark companions, but \gaia's astrometric precision (when combined with a mass estimate for the luminous star) will allow for precise mass measurements of the unseen companions.

Since longer period orbits subtend a larger angle on the sky, \gaia's sensitivity improves with orbital period up to orbits with periods as long as the mission lifetime. Therefore, a five year extension to \gaia's lifetime will not only allow for longer period orbits to be measured, but these binaries will also be the most precisely measured.

\acknowledgements
The authors thank the anonymous referee for comments which greatly improved the quality of the manuscript. The authors additionally acknowledge useful conversations with Kelle Cruz and Daniella Bardalez Gagliuffi. The authors thank J.\ Thorstensen for sharing a version of his code, SKYCALC. Furthermore, the authors are grateful for fruitful discussions with Berry Holl about the astrometric precision of \gaia\ observations. J.J.A.\ acknowledges support by the Danish National Research Foundation (DNRF132). The simulations within this work were run on the Metropolis HPC Facility at the CCQCN Center of the University of Crete, supported by the European Union Seventh Framework Programme (FP7-REGPOT-2012-2013-1) under grant agreement no.\ 316165. It is with great appreciation that K.B.\ and J.J.A.\ thank the hospitality of the Tata Institute for Fundamental Research, where the majority of this work was undertaken. K.B. is grateful for support from the Jeffrey L. Bishop Fellowship.

\software{{\tt astropy} \citep{astropy}, {\tt SKYCALC} (J.\ Thorstensen, private communication), {\tt emcee} \citep{Foreman-Mackey2013}, {\tt NumPy} \citep{numpy}, {\tt SciPy} \citep{scipy}, {\tt matplotlib} \citep{matplotlib}}

\bibliographystyle{aasjournal}
\bibliography{gaia}

\end{document}